# Beyond Accuracy: Benchmarking Uncertainty Quantification in Retinal Disease Screening

Mahvash Jebeli[1,*] Mohammad Eslami[2,*] Nazlee Zebardast[2] Lucia Sobrin[2] Tobias Elze[2]
[1]Independent Researcher [2]Harvard Medical School
moha_esla@meei.harvard.edu
[*]Equal contribution

## Abstract

*Reliable uncertainty quantification (UQ) is essential for evaluating retinal AI systems beyond classification accuracy, yet it remains unclear whether existing UQ strategies generalize across diseases, model architectures, and reliability criteria. We introduce a large-scale, multi-disease benchmark comprising over $242,000$ fundus photographs from $32$ datasets covering diabetic retinopathy, age-related macular degeneration, and glaucoma. We evaluate six representative UQ strategies across three Vision Transformer configurations, assessing discrimination, probabilistic calibration, selective prediction, and conformal prediction-set coverage. Our findings reveal no universally superior model or UQ strategy: methods that improve reliability in one disease or architecture may underperform in another, while improvements in discrimination, calibration, and selective prediction do not necessarily align. Notably, computationally intensive methods do not consistently outperform simple confidence scores, and domain-specific foundation models and calibration-oriented loss functions do not guarantee greater reliability. In contrast, conformal prediction provides a complementary framework for achieving near-nominal marginal prediction-set coverage across diverse settings, subject to its statistical assumptions. These findings challenge universal UQ recommendations and highlight the need for disease-specific, model-specific, and multidimensional reliability evaluation before clinical deployment. We release standardized dataset splits, trained checkpoints, evaluation code, and an interactive research demonstration to support reproducible benchmarking.*

## 1. Introduction

The application of Artificial Intelligence (AI) in healthcare holds transformative clinical and societal importance, particularly for large-scale preventative screening and automated medical image analysis in underserved populations [31, 33, 57]. Retinal fundus photography serves as a primary, non-invasive window into sight-threatening and systemic pathologies—most notably diabetic retinopathy (DR), age-related macular degeneration (AMD), and glaucoma, which collectively represent major causes of vision loss worldwide. While several AI-based retinal screening systems have achieved regulatory milestones (including FDA clearance for Digital Diagnostics' IDx-DR, EyeNuk's EyeArt, AEYE Health, and iPredict for AMD, alongside Verily's CE-marked DR system in India), widespread clinical adoption remains constrained by fundamental concerns regarding model trustworthiness, probabilistic miscalibration, and silent failures on ambiguous cases [14, 54, 61, 67].

In real-world medical screening, a deep learning classifier cannot act as an infallible oracle. Inherent data noise (e.g., media opacity, poor illumination, ungradable acquisition artifacts, or subtle early-stage lesions) and model limitations inevitably introduce predictive uncertainty [6, 25, 41]. Consequently, evaluating Uncertainty Quantification (UQ) is an important component of assessing the reliability of machine-assisted decision-making [10, 73]. When a screening model encounters an ambiguous examination, a reliable UQ mechanism can support selective prediction, allowing the system to abstain from automated diagnosis and refer uncertain cases for ophthalmologist review—a potential "learning to defer" or "second opinion" human-in-the-loop workflow [7, 13, 34]. However, the effectiveness of such strategies must be systematically evaluated before clinical deployment.

Despite growing interest in medical UQ across radiology [54], histopathology [18], and dermatology [71], systematic benchmarking in ophthalmic image analysis remains fragmented. Prior work has established the importance of evaluating reliability beyond predictive accuracy. Band et al. [9] benchmarked Bayesian UQ methods for diabetic retinopathy (DR), while Google's Plex framework [63] evaluated pretrained model reliability across diverse vision and language tasks, including DR under distribution shifts. More recently, Kim et al. [32] examined clinical reliability in DR screening, including calibration, clinician–model uncertainty agreement, acquisition and device robustness, and model-update consistency. However, these studies did not provide a standardized multi-disease retinal benchmark systematically comparing representative UQ strategies across discrimination, probabilistic calibration, and selective prediction. Similarly, glaucoma-focused benchmarks such as AIROGS [16] and uncertainty-aware approaches such as UIOS [66] and Bayesian DenseNet [5] address complementary aspects of reliability but do not establish such a comparative evaluation framework across modern self-supervised Vision Transformers and retinal foundation models.

To address this gap, we construct and release a large-scale multi-disease retinal screening benchmark for uncertainty quantification, comprising over $242,000$ fundus photographs compiled from more than 30 international cohorts across DR ($> 100,000$), AMD ($> 28,000$), and glaucoma ($> 114,000$). Rather than proposing a bespoke training architecture or exhaustively comparing UQ algorithms, we evaluate representative, widely used uncertainty estimation strategies with different implementation requirements, ranging from single-pass probability scores to stochastic inference, lightweight ensembles, and conformal prediction. We take the practical perspective of clinical AI evaluators and system builders: **how reliably do existing UQ strategies support probability calibration and identification of uncertain predictions across retinal diseases and pretrained Vision Transformer architectures?** Our benchmark provides a standardized framework for evaluating discrimination, calibration, and selective prediction as complementary dimensions of model reliability.

**Summary of Contributions:**

1. **A 242,000+ Image Multi-Disease UQ Benchmark:** We curate and standardize 32 global fundus datasets across three major blinding diseases (DR, AMD, and glaucoma), with stratified train, validation, test, and held-out 20% calibration splits to support fair, reproducible evaluation of uncertainty estimation and conformal prediction.
2. **Standardized Reliability Evaluation Across Architectures & Loss Formulations:** We benchmark six representative UQ strategies (MSP, Entropy, MC Dropout, TTA, Deep Ensembles [Std, Total, Aleatoric, and Epistemic], and Split Conformal Prediction) using complementary measures of discrimination, probabilistic calibration, and selective prediction. We evaluate general-domain self-supervised `ViT` (DINOv2) [49], retinal foundation model `RETFound-Green` [19], and calibration-oriented sample-dependent focal loss `ViT(FLSD-53)` [44].
3. **Identifying Disease- and Architecture-Dependent UQ Limitations:** We demonstrate that `ViT(FLSD-53)`—originally validated on natural image benchmarks—can exhibit substantial miscalibration on imbalanced binary retinal screening tasks. Furthermore, we show that TTA can improve calibration for DINOv2 `ViT` while degrading `RETFound-Green` performance on AMD, highlighting the importance of disease- and architecture-specific UQ evaluation.
4. **Open-Source Benchmark & Interactive Evaluation Tool:** We release standardized CSV splits, preprocessing pipelines, trained model weights, evaluation scripts, and a Hugging Face Spaces research demonstration that enables researchers and clinical AI evaluators to inspect per-image predictions and uncertainty profiles across models.

# 2. Benchmark Construction and Methodology

## 2.1. Global Multi-Disease Dataset Compilation

To establish a rigorous, large-scale benchmark representative of global screening populations, we aggregated 32 publicly available fundus photograph datasets containing diagnostic or referable grading labels for diabetic retinopathy (DR), age-related macular degeneration (AMD), and glaucoma (complete cohort breakdown in Appendix Table 3). This compilation yields over $100,000$ images for DR, $28,000$ for AMD, and $114,000$ for glaucoma. All diagnostic labels are harmonized into binary screening targets ($0 =$ healthy / non-referable, $1 =$ referable disease).

To prevent data leakage and preserve cross-study comparability, only samples explicitly designated as test sets by the original dataset publishers are included in our evaluation pool; all other samples are assigned to the development set (75% training, 25% validation). Whenever patient identifiers were available, we ensured patient-level separation during dataset construction. However, for several derived public datasets, cross-dataset identifiers were unavailable. Therefore, we performed a comprehensive image-level leakage audit across all training and test partitions using DINOv2 image embeddings with cosine similarity greater than 0.98, followed by manual inspection of all high-confidence candidate pairs. Furthermore, because post-hoc calibration and Split Conformal Prediction require a dedicated calibration cohort, we labeled a stratified 20% random subset from the test pool to serve strictly as a held-out calibration set ($\mathcal{D}_{\text{cal}}$), which is excluded from all final test metrics ($\mathcal{D}_{\text{test}}$). This split design mirrors realistic clinical validation where evaluators calibrate a pre-trained model on a local reference sample before auditing test performance.

## 2.2. Preprocessing and Vision Transformer Backbones

All fundus photographs undergo automated region-of-interest (ROI) detection, circular retinal cropping, and zero-padded square resizing ($392 \times 392$ pixels) via the open-source EyeQ pipeline [20, 21], followed by ImageNet mean/standard deviation normalization. During training, we apply stochastic data augmentations including resizing to 110% followed by random cropping, random horizontal/vertical flips, rotations ($\leq 10°$), color jitter (brightness, contrast, saturation, hue), and Gaussian blur.

We evaluate three model configurations based on two pretrained backbones:

- **`ViT` (DINOv2):** A general-domain Vision Transformer pretrained via self-supervised learning on ImageNet [49].
- **`RETFound-Green`:** A domain-specific ophthalmic foundation model sharing the same ViT architecture but pretrained on 1.6 million unlabeled retinal fundus photographs [19].
- **`ViT(FLSD-53)`:** An uncertainty- and calibration-oriented baseline trained with the sample-dependent focal loss of Mukhoti et al. [44], which applies a higher focusing parameter ($\gamma = 5$) to hard, low-confidence training examples ($\hat{p}_y \in [0, 0.2)$) and a lower focusing parameter ($\gamma = 3$) to easier examples ($\hat{p}_y \in [0.2, 1]$).

For all three backbones, pretrained feature extraction layers are frozen and a custom linear classification head (2 output logits, 10% dropout rate) is trained for up to 150 epochs using Adam (lr $= 5 \times 10^{-4}$, batch size 250, `ReduceLROnPlateau` patience 10, early stopping patience 15 on validation loss) with inverse class-frequency weighting to mitigate class imbalance.

### 2.3. Uncertainty Quantification Strategies

We evaluate six UQ strategies alongside a single-pass reference baseline (Table 1; extended taxonomy in Appendix Section A.2):

1. **Baseline (Reference):** We retain the original single-pass model predictions without applying any UQ method. For selective prediction, we use $1 - p(y = \textit{disease} \mid x)$ to construct a reference risk–coverage curve based on the predicted disease probability. This baseline serves only as a comparison reference and is not considered an uncertainty estimation method.
2. **Maximum Softmax Probability (MSP):** Uses a single deterministic forward pass, defining uncertainty as $u_{\text{MSP}}(x) = 1 - \max_c p(y = c \mid x)$.
3. **Predictive Entropy:** Measures uncertainty from a single forward pass using $H(p) = -\sum_c p_c \log p_c$.
4. **Monte Carlo (MC) Dropout:** Performs $T = 50$ stochastic forward passes with dropout enabled at inference [23]. Predictions are averaged for classification, while classwise standard deviation quantifies predictive variability. We use more samples than prior benchmarks ($T = 5$ in [9]; $T = 10$ in [23]) to reduce Monte Carlo estimation error.
5. **Test-Time Augmentation (TTA):** Generates $K = 20$ stochastic augmented views per image, averages their softmax predictions, and measures uncertainty using classwise standard deviation [43, 47].
6. **Deep Ensembles:** We independently train $N = 5$ classification heads with different random initializations and train/validation splits, keeping the pretrained feature extractor frozen [36, 51]. We average their predictive probabilities and quantify uncertainty using classwise standard deviation (**Std**), total predictive entropy (**Total**), expected member entropy (**Aleatoric**), and mutual information (**Epistemic**). The entropy-based terms are

$$H(\bar{p}), \quad \frac{1}{N}\sum_{n=1}^{N} H(p_n), \quad H(\bar{p}) - \frac{1}{N}\sum_{n=1}^{N} H(p_n),$$

respectively, where $\bar{p} = \frac{1}{N}\sum_n p_n$.
7. **Split Conformal Prediction (CP):** Wraps any base predictor using non-conformity scores calibrated on $\mathcal{D}_{\text{cal}}$ to construct statistically valid prediction sets $\mathcal{C}_\alpha(x)$ with marginal coverage guarantee $1 - \alpha$.

### 2.4. Evaluation Metrics

We evaluated model reliability across three complementary dimensions: discrimination, probabilistic calibration, and selective prediction. Discriminative performance was assessed using the area under the receiver operating characteristic curve (AUROC) and area under the precision–recall curve (AUPRC). Calibration was evaluated using expected calibration error (ECE; 10 bins), negative log-likelihood (NLL), and Brier score, measuring the agreement between predicted probabilities and observed outcomes. Selective prediction was assessed using the area under the risk–coverage curve (AURC), classification risk at 90% coverage (Risk@90% Coverage), and maximum coverage at an empirical 5% classification risk threshold (Coverage@5% Risk). Here, coverage represents the proportion of predictions retained for automated classification, while risk denotes the classification error rate among retained predictions. Lower AURC and risk, and higher coverage at a fixed risk threshold, indicate better selective prediction performance. To indicate the preferred direction of each metric, ↑ denotes higher values are better and ↓ denotes that lower values are better.

To assess statistical uncertainty, we computed 95% confidence intervals using 1,000 stratified bootstrap resamples, preserving the original class distribution. Paired resampling was used to estimate confidence intervals for differences between UQ methods, with conformal calibration samples excluded from evaluation.

## 3. Experimental Results

### 3.1. Disease Detection and Probabilistic Calibration Analysis

We first evaluate how uncertainty estimation strategies affect discriminative classification (Figure 1 and Table 2, top) and probabilistic calibration (Figure 2 and Table 2, bottom).

Table 1. Comparison of evaluated uncertainty quantification (UQ) strategies and their implementation requirements. The baseline uses single-pass predictions without UQ. Split conformal prediction (CP) can be applied to the predictive probabilities produced by the other methods.

| | Baseline | MSP / Entropy | TTA | MC Dropout | CP | Deep Ensemble | | | |
|---|---|---|---|---|---|---|---|---|---|
| | | | | | | Std | Aleatoric | Epistemic | Total |
| Prediction | Single-pass probabilities | Single-pass probabilities | Mean over augmentations | Mean over stochastic passes | Prediction set | Mean over ensemble heads | Same | Same | Same |
| Uncertainty / output | – | $1 - \max_c p_c$ or $H(p)$ | Predictive Std | Predictive Std | Set size (informativeness) | Predictive Std | Expected entropy | Mutual information | Predictive entropy |
| Additional requirements | None | None | Multiple augmented views | Dropout-enabled inference | Held-out calibration set | Multiple trained heads | Same | Same | Same |
| Inference cost | Single pass | Single pass | $K = 20$ passes | $T = 50$ passes | Set construction after calibration | $N = 5$ head evaluations | Same | Same | Same |

Paired bootstrap confidence intervals are used to assess the robustness of observed differences.

1. **Discriminative Performance Across Diseases and Backbones:** Baseline AUROC ranged from 0.895–0.913 for glaucoma, 0.883–0.912 for AMD, and 0.792–0.819 for DR. However, precision–recall analysis revealed substantial disease-dependent differences. Baseline AUPRC ranged from 0.804–0.819 for glaucoma and 0.799–0.805 for DR, compared with 0.421–0.540 for AMD. These findings highlight the importance of reporting AUPRC alongside AUROC, particularly when evaluating datasets with different class distributions.
2. **Divergent Impact of Test-Time Augmentation (TTA):** For the general-domain `ViT` (DINOv2) backbone, averaging predictions across $K = 20$ test-time augmentations improved DR AUROC from 0.792 to 0.806 (paired difference: $+0.014$, 95% CI: $[+0.009, +0.021]$) and glaucoma AUROC from 0.908 to 0.922. However, AMD AUROC decreased slightly from 0.911 to 0.903. MC Dropout and Deep Ensembles generally produced smaller changes in discriminative performance relative to the single-pass baseline. In contrast, `RETFound-Green`, which achieved the highest baseline DR AUROC (0.819), exhibited substantial sensitivity to TTA. In AMD, its AUROC decreased from 0.883 to 0.810 (paired difference: $-0.072$, 95% CI: $[-0.107, -0.040]$), while AUPRC declined from 0.421 to 0.266. These results indicate that the effects of test-time augmentation depend strongly on the disease and pretrained representation.
3. **Calibration Efficacy and Limitations of Focal Loss (`ViT(FLSD-53)`):** As shown in the reliability diagrams (Figure 2) and Table 2 (bottom), calibration performance varied substantially across diseases and architectures. TTA improved calibration for general-domain `ViT` in AMD, reducing ECE from 0.257 to 0.183 and Brier score from 0.156 to 0.112, and in glaucoma, reducing ECE from 0.171 to 0.059 and NLL from 0.413 to 0.317. However, TTA increased ViT ECE in DR from 0.045 to 0.063, demonstrating that calibration improvements are not universal. Notably, `ViT(FLSD-53)` exhibited higher baseline ECE than the cross-entropy-trained configurations across all three diseases: 0.350 in AMD, 0.165 in DR, and 0.277 in glaucoma. These findings suggest that calibration-oriented loss functions developed for natural image classification do not necessarily transfer to binary retinal screening without task-specific adaptation.

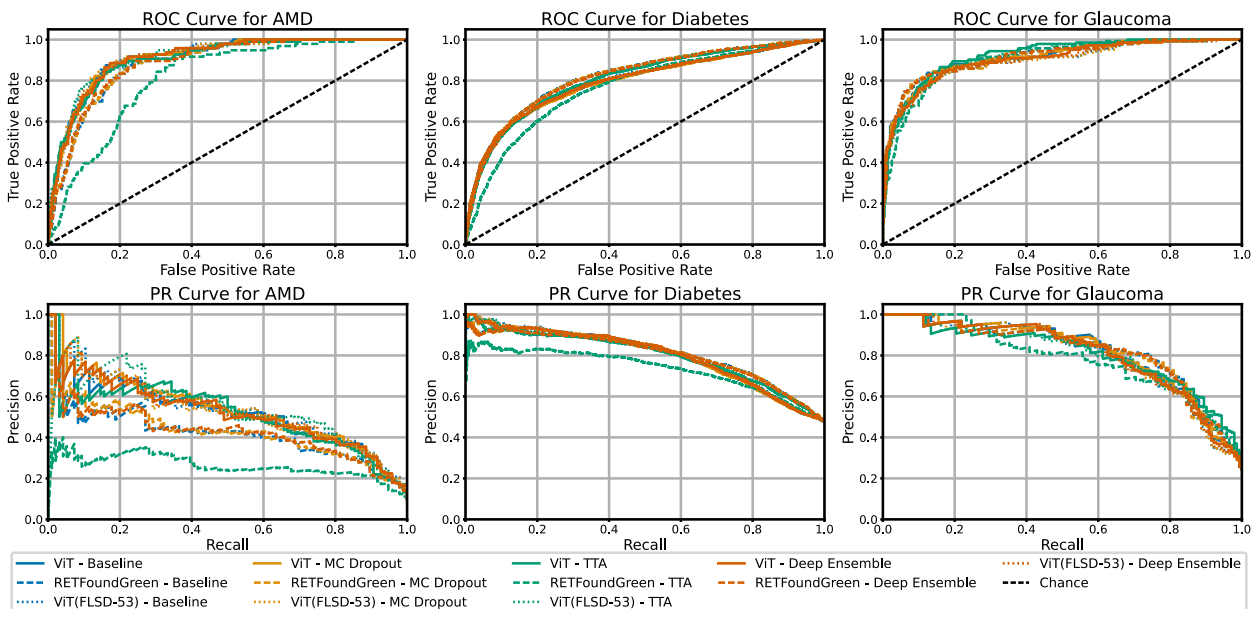

Figure 1. Receiver Operating Characteristic (ROC) and Precision–Recall (PR) curves across AMD, Diabetic Retinopathy, and Glaucoma for `ViT`, `RETFound-Green`, and `ViT(FLSD-53)` under different inference-time UQ strategies. Best viewed at 200% magnification. Also available in the Appendix.

## 3.2. Selective Prediction and Risk–Coverage Trade-offs

To evaluate how effectively uncertainty scores rank sample reliability for selective prediction and potential human-in-the-loop abstention, Figure 3 and Table 5 (appendix) report risk–coverage performance across models and diseases, while Figure 8(appendix) visualizes the distributions of uncertainty scores. Paired bootstrap comparisons further assess differences between methods.

- **Selective Prediction Gains in Glaucoma:** Glaucoma benefited substantially from uncertainty-guided selective prediction across multiple architectures. For `ViT`, TTA achieved the lowest AURC (0.042), compared with 0.078 for MSP and 0.106 for Deep Ensemble predictive standard deviation. The improvement over MSP was sup-

Table 2. Comprehensive Benchmark Results for Discriminative Detection (AUROC ↑, AUPRC ↑; top table) and Probabilistic Calibration (NLL ↓, ECE ↓, Brier Score ↓; bottom table) across AMD, Diabetic Retinopathy, and Glaucoma.

| | ViT | | | | RETFoundGreen | | | | ViT(FLSD-53) | | | |
|---|---|---|---|---|---|---|---|---|---|---|---|---|
| Disease | Baseline | MC Dropout | TTA | Deep Ensemble | Baseline | MC Dropout | TTA | Deep Ensemble | Baseline | MC Dropout | TTA | Deep Ensemble |
| **AUROC** | | | | | | | | | | | | |
| AMD | 0.911 [0.882, 0.934] | 0.911 [0.882, 0.935] | 0.903 [0.874, 0.929] | 0.910 [0.880, 0.933] | 0.883 [0.852, 0.910] | 0.884 [0.854, 0.912] | 0.810 [0.772, 0.845] | 0.884 [0.853, 0.912] | **0.912 [0.885, 0.936]** | 0.911 [0.884, 0.935] | 0.911 [0.883, 0.935] | 0.911 [0.885, 0.935] |
| Diabetes | 0.792 [0.777, 0.804] | 0.791 [0.776, 0.803] | 0.806 [0.793, 0.818] | 0.794 [0.780, 0.806] | **0.819 [0.806, 0.830]** | 0.818 [0.806, 0.830] | 0.766 [0.752, 0.778] | 0.816 [0.804, 0.827] | 0.796 [0.783, 0.809] | 0.796 [0.782, 0.809] | 0.809 [0.796, 0.821] | 0.795 [0.782, 0.808] |
| Glaucoma | 0.908 [0.878, 0.938] | 0.909 [0.880, 0.938] | **0.922 [0.897, 0.946]** | 0.902 [0.870, 0.934] | 0.913 [0.882, 0.941] | 0.911 [0.880, 0.941] | 0.905 [0.878, 0.930] | 0.913 [0.883, 0.941] | 0.895 [0.861, 0.925] | 0.894 [0.860, 0.925] | 0.913 [0.886, 0.937] | 0.896 [0.863, 0.926] |
| **AUPRC** | | | | | | | | | | | | |
| AMD | 0.525 [0.435, 0.621] | 0.541 [0.450, 0.636] | 0.514 [0.426, 0.614] | 0.525 [0.435, 0.621] | 0.421 [0.348, 0.519] | 0.432 [0.357, 0.533] | 0.266 [0.223, 0.343] | 0.432 [0.355, 0.527] | 0.540 [0.464, 0.634] | 0.534 [0.457, 0.631] | **0.543 [0.459, 0.652]** | 0.534 [0.457, 0.627] |
| Diabetes | 0.799 [0.784, 0.812] | 0.798 [0.783, 0.812] | 0.797 [0.782, 0.811] | 0.801 [0.786, 0.814] | **0.805 [0.789, 0.819]** | 0.804 [0.788, 0.819] | 0.735 [0.717, 0.752] | 0.803 [0.786, 0.817] | 0.803 [0.789, 0.817] | 0.803 [0.789, 0.817] | 0.803 [0.789, 0.817] | 0.801 [0.787, 0.815] |
| Glaucoma | 0.817 [0.760, 0.871] | 0.817 [0.763, 0.870] | 0.808 [0.750, 0.867] | 0.808 [0.753, 0.863] | 0.819 [0.761, 0.871] | **0.820 [0.763, 0.871]** | 0.784 [0.733, 0.839] | 0.818 [0.760, 0.870] | 0.804 [0.749, 0.859] | 0.804 [0.747, 0.859] | 0.802 [0.743, 0.855] | 0.805 [0.750, 0.858] |

| | ViT | | | | RETFoundGreen | | | | ViT(FLSD-53) | | | |
|---|---|---|---|---|---|---|---|---|---|---|---|---|
| Disease | Baseline | MC Dropout | TTA | Deep Ens. | Baseline | MC Dropout | TTA | Deep Ens. | Baseline | MC Dropout | TTA | Deep Ens. |
| **NLL** | | | | | | | | | | | | |
| AMD | 0.490 [0.448, 0.535] | 0.484 [0.447, 0.524] | **0.357 [0.329, 0.387]** | 0.514 [0.472, 0.561] | 0.536 [0.494, 0.581] | 0.528 [0.489, 0.570] | 0.383 [0.361, 0.405] | 0.549 [0.507, 0.594] | 0.569 [0.557, 0.580] | 0.570 [0.559, 0.581] | 0.548 [0.537, 0.559] | 0.570 [0.558, 0.581] |
| Diabetes | 0.547 [0.533, 0.563] | 0.548 [0.534, 0.562] | 0.549 [0.535, 0.563] | 0.544 [0.530, 0.559] | **0.530 [0.518, 0.543]** | 0.531 [0.519, 0.544] | 0.586 [0.576, 0.598] | 0.531 [0.519, 0.545] | 0.617 [0.612, 0.621] | 0.617 [0.613, 0.621] | 0.620 [0.616, 0.624] | 0.614 [0.610, 0.619] |
| Glaucoma | 0.413 [0.374, 0.453] | 0.417 [0.379, 0.453] | **0.317 [0.285, 0.349]** | 0.412 [0.372, 0.452] | 0.368 [0.336, 0.401] | 0.374 [0.344, 0.405] | 0.358 [0.331, 0.385] | 0.373 [0.342, 0.407] | 0.554 [0.542, 0.567] | 0.555 [0.543, 0.568] | 0.526 [0.516, 0.538] | 0.566 [0.555, 0.579] |
| **ECE** | | | | | | | | | | | | |
| AMD | 0.257 [0.241, 0.274] | 0.264 [0.248, 0.279] | 0.183 [0.170, 0.198] | 0.270 [0.253, 0.286] | 0.276 [0.260, 0.292] | 0.282 [0.266, 0.297] | **0.173 [0.162, 0.184]** | 0.288 [0.272, 0.304] | 0.350 [0.344, 0.358] | 0.350 [0.345, 0.359] | 0.336 [0.329, 0.343] | 0.350 [0.345, 0.359] |
| Diabetes | 0.045 [0.038, 0.057] | 0.044 [0.038, 0.057] | 0.063 [0.057, 0.070] | 0.044 [0.036, 0.055] | 0.050 [0.042, 0.062] | 0.049 [0.043, 0.062] | 0.052 [0.040, 0.065] | **0.043 [0.037, 0.056]** | 0.165 [0.153, 0.176] | 0.164 [0.152, 0.176] | 0.173 [0.161, 0.185] | 0.164 [0.152, 0.176] |
| Glaucoma | 0.171 [0.154, 0.190] | 0.175 [0.161, 0.195] | **0.059 [0.052, 0.088]** | 0.164 [0.146, 0.183] | 0.127 [0.117, 0.151] | 0.138 [0.125, 0.160] | 0.086 [0.065, 0.110] | 0.132 [0.122, 0.156] | 0.277 [0.257, 0.297] | 0.286 [0.265, 0.306] | 0.261 [0.234, 0.286] | 0.294 [0.274, 0.314] |
| **Brier Score** | | | | | | | | | | | | |
| AMD | 0.156 [0.142, 0.171] | 0.155 [0.142, 0.170] | **0.112 [0.100, 0.123]** | 0.164 [0.150, 0.180] | 0.172 [0.157, 0.187] | 0.171 [0.157, 0.185] | 0.118 [0.109, 0.127] | 0.177 [0.162, 0.192] | 0.190 [0.185, 0.196] | 0.191 [0.185, 0.196] | 0.180 [0.175, 0.186] | 0.190 [0.185, 0.196] |
| Diabetes | 0.183 [0.177, 0.189] | 0.183 [0.178, 0.189] | 0.182 [0.177, 0.188] | 0.182 [0.176, 0.188] | **0.175 [0.170, 0.180]** | **0.175 [0.171, 0.181]** | 0.199 [0.194, 0.204] | **0.175 [0.171, 0.181]** | 0.213 [0.211, 0.215] | 0.213 [0.211, 0.215] | 0.214 [0.213, 0.216] | 0.212 [0.210, 0.214] |
| Glaucoma | 0.129 [0.113, 0.145] | 0.130 [0.115, 0.145] | **0.095 [0.082, 0.107]** | 0.128 [0.113, 0.144] | 0.109 [0.097, 0.123] | 0.111 [0.099, 0.124] | 0.112 [0.101, 0.123] | 0.111 [0.099, 0.125] | 0.183 [0.177, 0.189] | 0.183 [0.177, 0.189] | 0.169 [0.164, 0.175] | 0.188 [0.183, 0.194] |

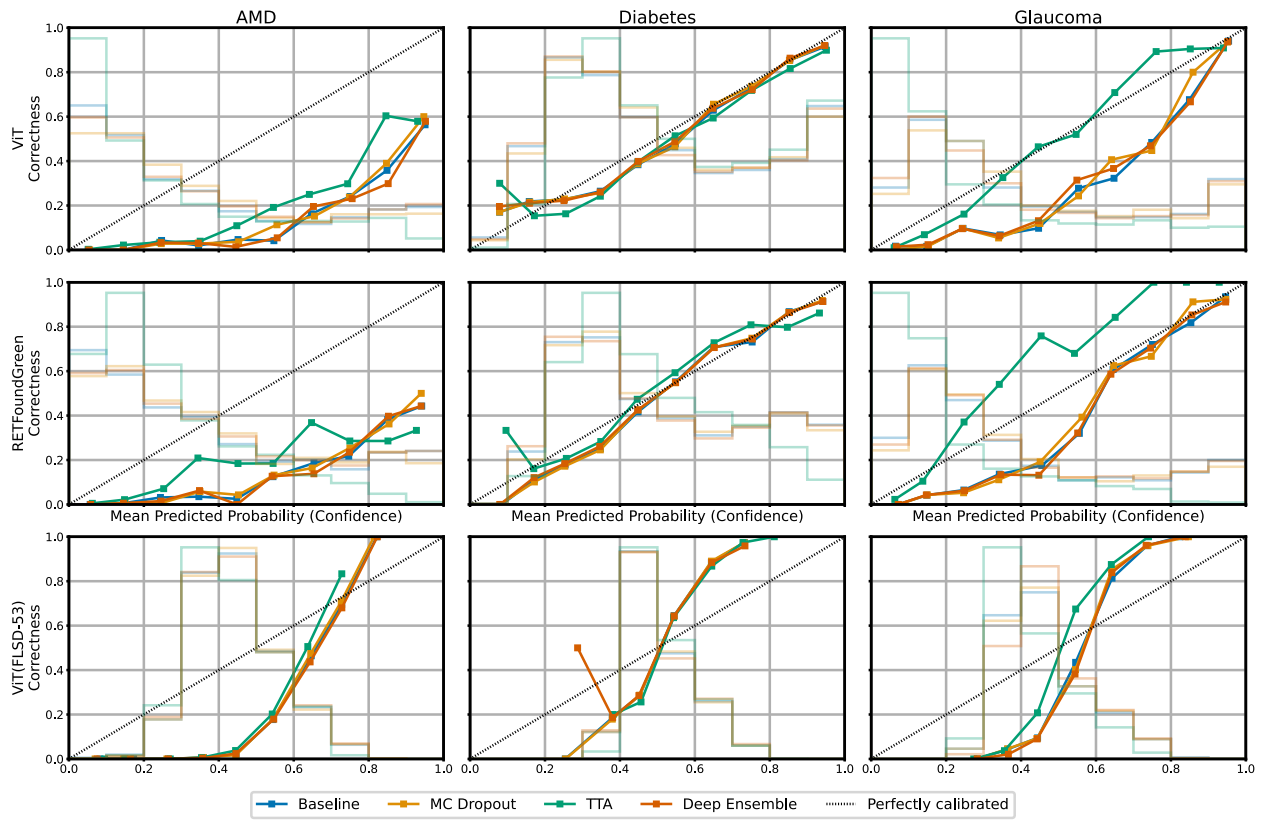

Figure 2. Reliability diagrams (observed empirical accuracy vs. predicted confidence) across diseases, architectures, and inference-time UQ strategies. Well-calibrated models track the dashed diagonal. Best viewed at 200% magnification. Also available in the Appendix.

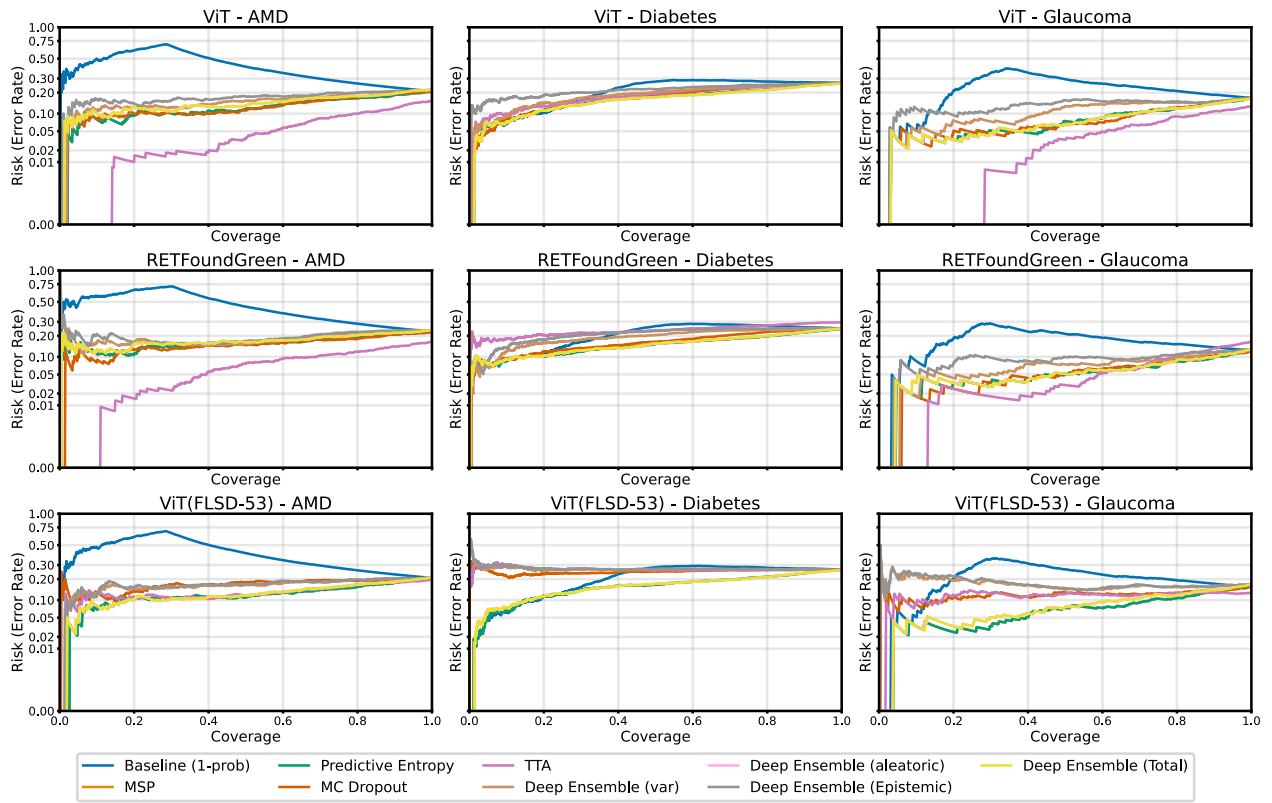

Figure 3. Risk–coverage curves across retinal diseases, model configurations, and uncertainty estimation methods. Coverage denotes the fraction of predictions retained after ranking samples by their respective scores, while risk represents the classification error rate among retained predictions. The baseline curve uses the original disease prediction score as a reference ranking rather than an uncertainty estimate. Lower curves indicate better selective prediction performance. Best viewed at 200% magnification. Also available in the Appendix.

ported by paired bootstrap analysis (AURC difference: $-0.036$, 95% CI: $[-0.054, -0.019]$). TTA also reduced Risk@90% Coverage from 14.5% (MSP) to 10.8% and achieved 60.8% empirical Coverage@5% Risk, compared with 43.7% for MSP and 17.5% for Deep Ensemble standard deviation. For `RETFound-Green`, TTA achieved an AURC of 0.053 and Coverage@5% Risk of 58.9%. These results demonstrate the potential of uncertainty-guided abstention to reduce classification errors among retained predictions, although the achievable coverage depends on the model and uncertainty strategy.

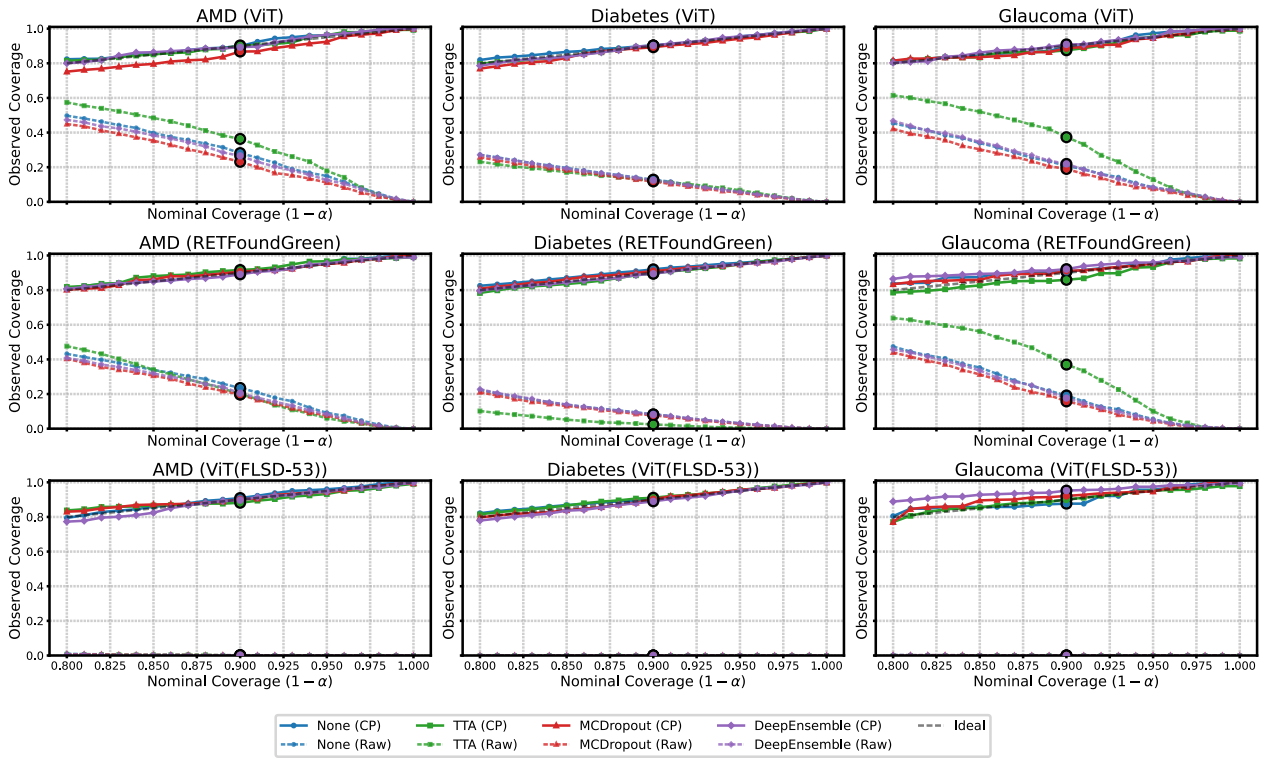

Figure 4. Observed versus nominal prediction-set coverage across diseases, models, and inference methods. Raw prediction sets exhibit substantial empirical undercoverage, whereas conformal prediction brings observed coverage closer to the nominal target. Coverage alone does not characterize probability calibration or prediction-set informativeness. Best viewed at 200% magnification. Also available in the Appendix.

- **Disease- and Architecture-Dependent Performance in AMD:** In AMD, TTA provided substantial selective prediction improvements for both `ViT` and `RETFound-Green`. For `ViT`, TTA reduced AURC from 0.126 (MSP) to 0.052 (paired difference: $-0.073$, 95% CI: $[-0.092, -0.056]$), reduced Risk@90% Coverage from 18.6% to 12.2%, and increased Coverage@5% Risk from 4.7% to 57.3%. Similarly, for `RETFound-Green`, TTA reduced AURC from 0.162 to 0.072 and increased Coverage@5% Risk from 0.3% to 39.3%. Importantly, these selective prediction improvements occurred despite a substantial decline in RETFound-Green AMD AUROC under TTA, illustrating that discrimination and uncertainty-based error ranking capture distinct aspects of model reliability.
- **Limited Selective Prediction Coverage in Diabetic Retinopathy:** In contrast to glaucoma and AMD, DR exhibited more limited improvements in selective prediction. For `ViT`, MSP achieved an AURC of 0.167, compared with 0.184 for TTA and 0.190 for Deep Ensemble standard deviation. The higher AURC under TTA relative to MSP was supported by paired bootstrap analysis (difference: $+0.017$, 95% CI: $[+0.009, +0.025]$). For `RETFound-Green`, MSP achieved an AURC of 0.153, compared with 0.236 for TTA. Across all three model configurations, Coverage@5% Risk remained low, with a maximum of 6.0% among the evaluated methods. These results suggest that uncertainty-guided abstention is less effective at retaining substantial automated coverage under a stringent empirical risk threshold for DR in the evaluated datasets.
- **Overall Method Comparison:** No uncertainty estimation strategy consistently dominated across diseases, architectures, and evaluation metrics. TTA achieved substantial selective prediction gains in selected AMD and glaucoma settings but degraded performance in several DR configurations. Notably, Deep Ensemble predictive standard deviation did not consistently outperform simpler uncertainty scores; its AURC was higher than that of MSP across all nine disease–model combinations, with paired bootstrap confidence intervals excluding zero. These findings emphasize that computationally intensive uncertainty estimation does not necessarily translate into improved selective prediction and support evaluating UQ strategies separately for each disease and model configuration.

### 3.3. Conformal Prediction

Conformal prediction (CP) was evaluated as an additional uncertainty quantification layer applied to the prediction scores of the baseline models and inference-time UQ methods. Figure 4 compares observed and nominal prediction-set coverage across diseases and model configurations. Ideally, observed coverage should approximate the nominal target, while substantial undercoverage indicates that the constructed prediction sets fail to attain the requested coverage on the evaluation data.

At a nominal coverage of 90% ($\alpha = 0.1$), CP achieved observed coverage ranging from 86.78% to 90.92% for `ViT`, 85.96% to 92.12% for `RETFound-Green`, and 87.67% to 95.21% for `ViT(FLSD-53)` across diseases and inference methods (Table 6). In contrast, the corresponding raw prediction-set coverage ranged from 11.80% to 37.33%, 2.43% to 36.99%, and 0% to 0.20%, respectively. The largest discrepancy was observed for `ViT(FLSD-53)`, for which CP brought empirical coverage close to the nominal target despite extremely low raw coverage.

These findings demonstrate that conformalization can substantially improve prediction-set coverage across model configurations and diseases, including settings with poor probabilistic calibration. Importantly, CP does not directly correct the underlying predicted probabilities; instead, under the exchangeability assumption, it provides a marginal coverage guarantee for prediction sets. Thus, CP offers a complementary reliability mechanism to discrimination, calibration, and selective prediction, although near-nominal coverage alone does not establish clinical utility or guarantee reliable predictions for individual patients.

## 4. Discussion and Recommendations

Our large-scale benchmark across $242,000+$ retinal images and three major ophthalmic diseases reveals a central finding: **there is no universally superior model configuration or uncertainty quantification (UQ) strategy across diseases** **and reliability metrics.** Importantly, the differences are not limited to predictive accuracy; discrimination, probabilistic calibration, and selective prediction can respond differently to the same UQ method. These findings have several implications for developing reliable clinical AI systems:

1. **No Universal Winner: Reliability Is Disease-, Model-, and Metric-Dependent.** No model configuration or UQ method consistently dominated across AMD, DR, and glaucoma. Methods that improved reliability in one setting sometimes degraded performance in another, and improvements in discrimination did not necessarily translate into better calibration or selective prediction. For example, TTA improved selective prediction for ViT in AMD and glaucoma but worsened AURC in DR. This heterogeneity challenges the assumption that UQ strategies validated on one disease or architecture can be directly transferred to another. **Clinical AI reliability must therefore be evaluated for each intended application, rather than inferred from model architecture or performance on another task.**
2. **Greater UQ Complexity Does Not Guarantee Greater Reliability.** Computationally intensive approaches did not consistently outperform simple, single-pass uncertainty scores. In particular, Deep Ensemble predictive standard deviation yielded higher AURC than MSP across all nine disease–model combinations, with paired bootstrap confidence intervals excluding zero. Conversely, TTA provided substantial improvements in selected AMD and glaucoma settings. These findings suggest that uncertainty estimation methods should be selected based on demonstrated reliability gains and computational requirements, rather than methodological complexity alone.
3. **Foundation Models and Calibration Losses Require Task-Specific Validation.** Despite strong baseline discrimination, `RETFound-Green` exhibited substantial AMD AUROC degradation under TTA ($0.883 \rightarrow 0.810$), even as its calibration and selective prediction improved. Similarly, `ViT(FLSD-53)` exhibited consistently higher calibration error than cross-entropy-trained configurations. These findings demonstrate that neither domain-specific pretraining nor calibration-oriented training objectives guarantee reliable predictions across clinical tasks. Augmentation strategies, loss functions, and uncertainty estimates require independent validation for each disease and model configuration.
4. **Conformal Prediction Provides a Complementary Coverage Guarantee.** Unlike the heterogeneous effects observed for other UQ strategies, conformal prediction (CP) consistently brought empirical prediction-set coverage closer to the nominal target across models, diseases, and inference methods. At 90% nominal coverage, CP achieved observed coverage of approximately 86%–95%, despite substantial undercoverage in the corresponding raw prediction sets. This finding highlights CP as a broadly applicable complement to conventional UQ methods. Under the exchangeability assumption, CP provides a finite-sample marginal coverage guarantee without requiring well-calibrated softmax probabilities. However, coverage alone does not establish clinical utility: prediction-set size, the frequency of ambiguous predictions, and performance across clinically relevant subgroups must also be considered. We therefore recommend incorporating conformal coverage and prediction-set efficiency into the evaluation of uncertainty-aware clinical AI systems.

Together, these findings motivate a shift from identifying a single best UQ method toward **multidimensional, disease-specific, and model-specific reliability evaluation**. For clinical AI researchers, strong classification performance or successful uncertainty estimation on one benchmark should not be interpreted as sufficient evidence of reliability in another clinical setting.

### 4.1. Interactive Open-Source Demonstration for Clinical Evaluators

To bridge the gap between static benchmark tables and real-world clinical inspection, we developed and deployed an interactive web application hosted on Hugging Face Spaces (Figure 5), alongside our complete open-source training and evaluation repository.

As illustrated in Figure 5, the online demonstration allows clinicians, regulators, and AI researchers to upload arbitrary fundus photographs and immediately inspect: (1) automated EyeQ quality/ROI cropping, (2) predicted disease probabilities across all three diseases and Vision Transformer backbones, and (3) per-sample uncertainty gauges and Conformal Prediction sets. In the displayed clinical example, even though the macular/foveal region is not clearly visible, the baseline AMD model outputs an overconfident referral probability; however, inspecting the multi-model ensemble disagreement and UQ indicators reveals substantial epistemic inconsistency across weight initializations, providing an intuitive visual safeguard against silent diagnostic failure.

### 4.2. Limitations

While our benchmark encompasses 32 global datasets and $> 242,000$ images, several limitations remain. First, we froze the pretrained Vision Transformer backbones to isolate inference-time UQ performance under standardized linear probing; end-to-end fine-tuning may alter the interplay between TTA and foundation model representations. Second, due to cloud memory limits on free community hosting tiers, the live Hugging Face Spaces demo executes single-model UQ methods (MSP, Entropy, MC Dropout, TTA, and CP) in real time while providing offline scripts for 5-model Deep Ensembles. Third, bootstrap confidence intervals quantify sampling variability conditional on the trained models and evaluation cohorts, rather than variability across independent training runs or external clinical sites. Furthermore, some public datasets lack complete patient identifiers, limiting our ability to account for within-patient dependence and potential cross-dataset overlap. Finally, although conformal prediction achieved near-nominal marginal coverage, we did not systematically evaluate prediction-set efficiency or subgroup-conditional coverage. These aspects require further investigation before clinical implementation.

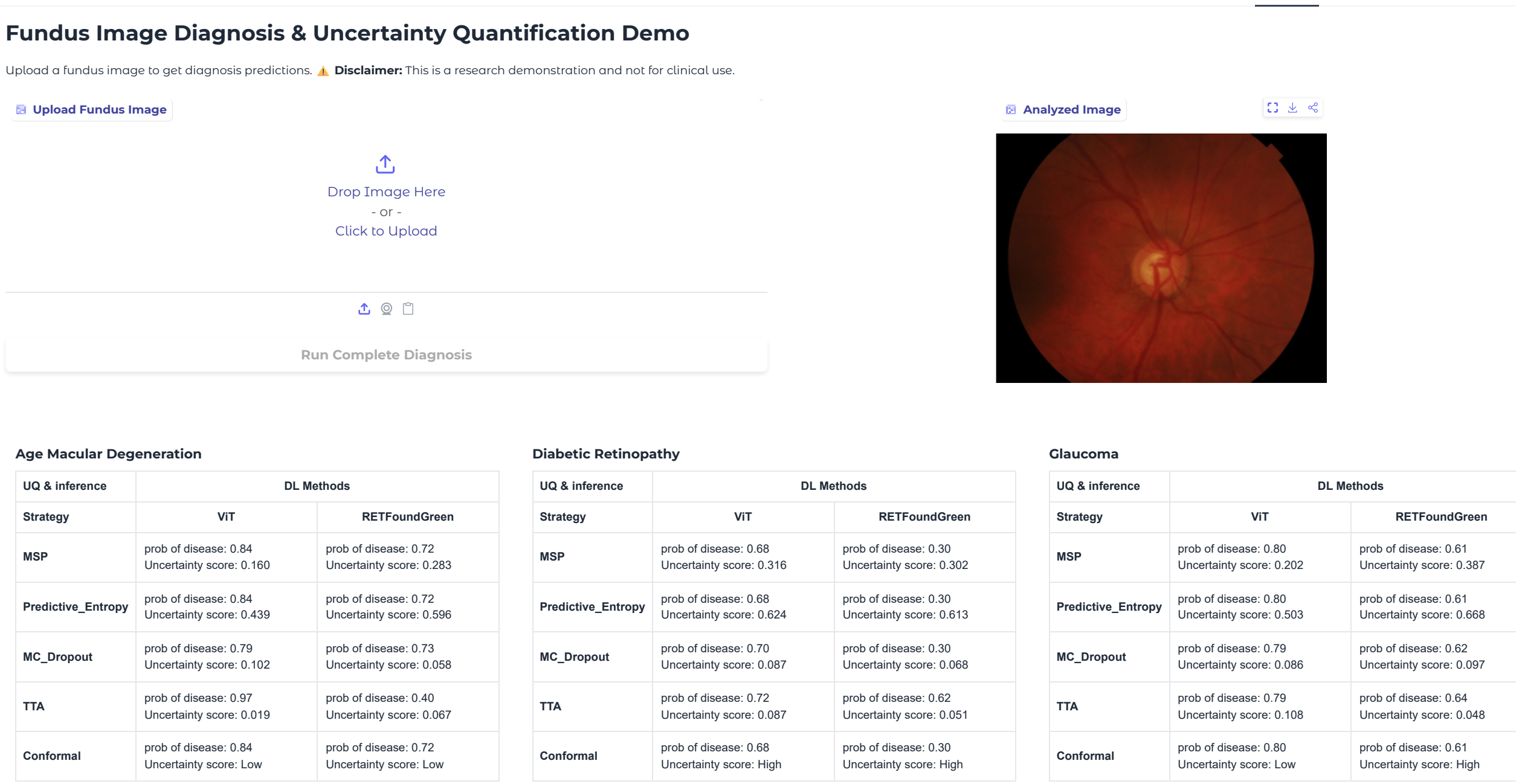

Figure 5. Snapshot of our deployed interactive clinical screening demonstration on Hugging Face Spaces. For any uploaded fundus photograph, the platform displays region-of-interest preprocessing, disease referral probabilities across trained backbones (`ViT`, `RETFound-Green`), and side-by-side inference-time UQ estimates (MSP, Entropy, MC Dropout, TTA, and Conformal Prediction sets). In this illustrative case, although the fovea is partially obscured, the AMD classifier issues an overconfident positive prediction, while cross-model disagreement and elevated uncertainty alert the clinician to inspect the examination manually.

## 5. Conclusion

We presented a large-scale, multi-disease benchmark of uncertainty quantification in retinal AI, evaluating multiple model configurations and UQ strategies across more than $242,000$ fundus photographs for AMD, diabetic retinopathy, and glaucoma. Our central finding is that **no model or UQ method consistently dominates across diseases, architectures, and reliability metrics**. More complex uncertainty methods do not necessarily outperform simpler alternatives, and improvements in discrimination, calibration, and selective prediction are not always aligned. In contrast, conformal prediction provides a broadly applicable framework for targeting marginal prediction-set coverage, although its clinical utility requires additional evaluation. These findings underscore the need for systematic, multidimensional reliability assessment tailored to each clinical application. By releasing standardized dataset splits, trained checkpoints, evaluation tools, and an interactive demonstration, we aim to facilitate reproducible evaluation and more informed development of uncertainty-aware retinal AI systems.

**LLM Usage Statement.** Large language models (LLMs) were used for proofreading, grammar correction, minor script development, and code review to identify potential errors and inconsistencies. No LLMs contributed to research ideation, experimental design, or substantive manuscript writing. All code and experimental results were reviewed and verified by the authors.

# A. Supplementary Material: Dataset Compilation, UQ Taxonomy, and Training Details

## A.1. Global 32-Dataset Benchmark Compilation

Table 3 lists all 32 datasets compiled for training, validation, calibration (20% stratified test subset), and testing across diabetic retinopathy, age-related macular degeneration, and glaucoma.

Table 3. List of used datasets in this study. The + shows the number of positive samples in that dataset.

| Dataset | Glaucoma | | DR | | AMD | | Test | Note |
|---|---|---|---|---|---|---|---|---|
| | Tot | + | Tot | + | Tot | + | | |
| AIROGS | 101120 | 3270 | | | | | | [16] |
| APTOS | | | 3662 | 1857 | | | | [29] |
| Aizawl | | | 495 | 452 | | | | [64] |
| BRSET | | | 16264 | 1070 | 16264 | 299 | | [24, 45, 46] |
| Cataract | 401 | 101 | | | | | | [70] |
| DDR | | | 13585 | 7328 | | | 4074 | [38] |
| DIARETDB1 | | | 89 | 89 | | | | [30] |
| DR1_DR2 | | | 1904 | 972 | | | | [55] |
| DeepDRiD | | | 1569 | 869 | | | 400 | [40] |
| Drishti | 101 | 70 | | | | | 51 | [59] |
| FIVES | 800 | 200 | 800 | 200 | 800 | 200 | 300 | [28] |
| G1020 | 1020 | 296 | | | | | | [8] |
| GRAPE | 631 | 631 | | | | | | [27] |
| IDRID | | | 516 | 348 | | | 103 | [56] |
| JICHI | | | 9939 | 3810 | | | | [62] |
| JSIEC | 51 | 13 | 144 | 106 | 38 | | | [12] |
| KCG | 1450 | 899 | | | | | | [60] |
| LES-AV | 11 | | | | | | | [48] |
| Mured | | | 1621 | 322 | 1621 | 131 | 332 | [58] |
| ODIR | 6985 | 326 | 6985 | 93 | 6985 | 280 | | [65] |
| ORIGA | 650 | 168 | | | | | | [72] |
| PAPILA | 488 | 155 | | | | | | [35] |
| RFMiD v1 (RIADD) | | | 2560 | 508 | 2560 | 138 | 640 | [52] |
| RFMiD v2 | | | 836 | 70 | 836 | 10 | 167 | [53] |
| MESSIDOR2 | | | 1748 | 731 | | | | [2] |
| SUSTech-SYSU | | | 1219 | 588 | | | | [39] |
| HYGD | 747 | 548 | | | | | | [3, 4] |
| TJDR | | | 257 | 257 | | | 55 | [42] |
| UNA-DR | | | 1437 | 726 | | | | [11] |
| e-ophtha | | | 237 | 121 | | | | [17] |
| eyePACS_orig | | | 35125 | 9315 | | | | [15] |
| iChallenge_ADAM | | | | | 400 | 89 | | [22] |
| iChallenge_GAMMA | 300 | 149 | | | | | 100 | [69] |
| iChallenge_REFUGE | 1200 | 120 | | | | | 400 | [50] |
| mBRSET | | | 4883 | 1134 | | | | [68] |

## A.2. Extended Taxonomy of Uncertainty Quantification Methods

[1, 26, 37]:

- **Deterministic Methods** produce uncertainty estimates through a single forward pass (MSP and Predictive Entropy). TTA acts as a deterministic/statistical hybrid when aggregating predictions across $K = 20$ stochastic input transformations.
- **Bayesian Approximation Methods** treat model weights stochastically; Monte Carlo Dropout ($T = 50$) approximates variational Bayesian inference in networks trained with dropout layers [23].
- **Statistical & Ensemble Methods** combine multiple independently trained predictors ($N = 5$ Deep Ensembles [36]) to approximate Bayesian model averaging and decompose predictive variance into aleatoric and epistemic components, while Split Conformal Prediction provides finite-sample, distribution-free coverage guarantees $(1 - \alpha)$ using a held-out calibration split ($\mathcal{D}_{\text{cal}}$).

Table 4. Comparison of Aleatoric and Epistemic Uncertainty Sources in Ophthalmic Screening.

| **Aleatoric Uncertainty** | **Epistemic Uncertainty** |
|---|---|
| Arises from inherent noise, randomness, or ambiguity in the data itself. | Arises from limitations in the model or insufficient training data. |
| Considered **irreducible**; cannot be reduced by collecting more of the same data. | Considered **reducible**; can be reduced with more diverse data or a better model. |
| **Sources in Fundus Imaging:** Sensor noise, motion artifacts, poor focus, early-stage pathology. | **Sources in Fundus Imaging:** Out-of-distribution data, lack of examples for rare diseases, model misspecification. |
| Quantifies the unpredictability of the system being measured. | Quantifies the model's lack of knowledge about the data-generating function. |

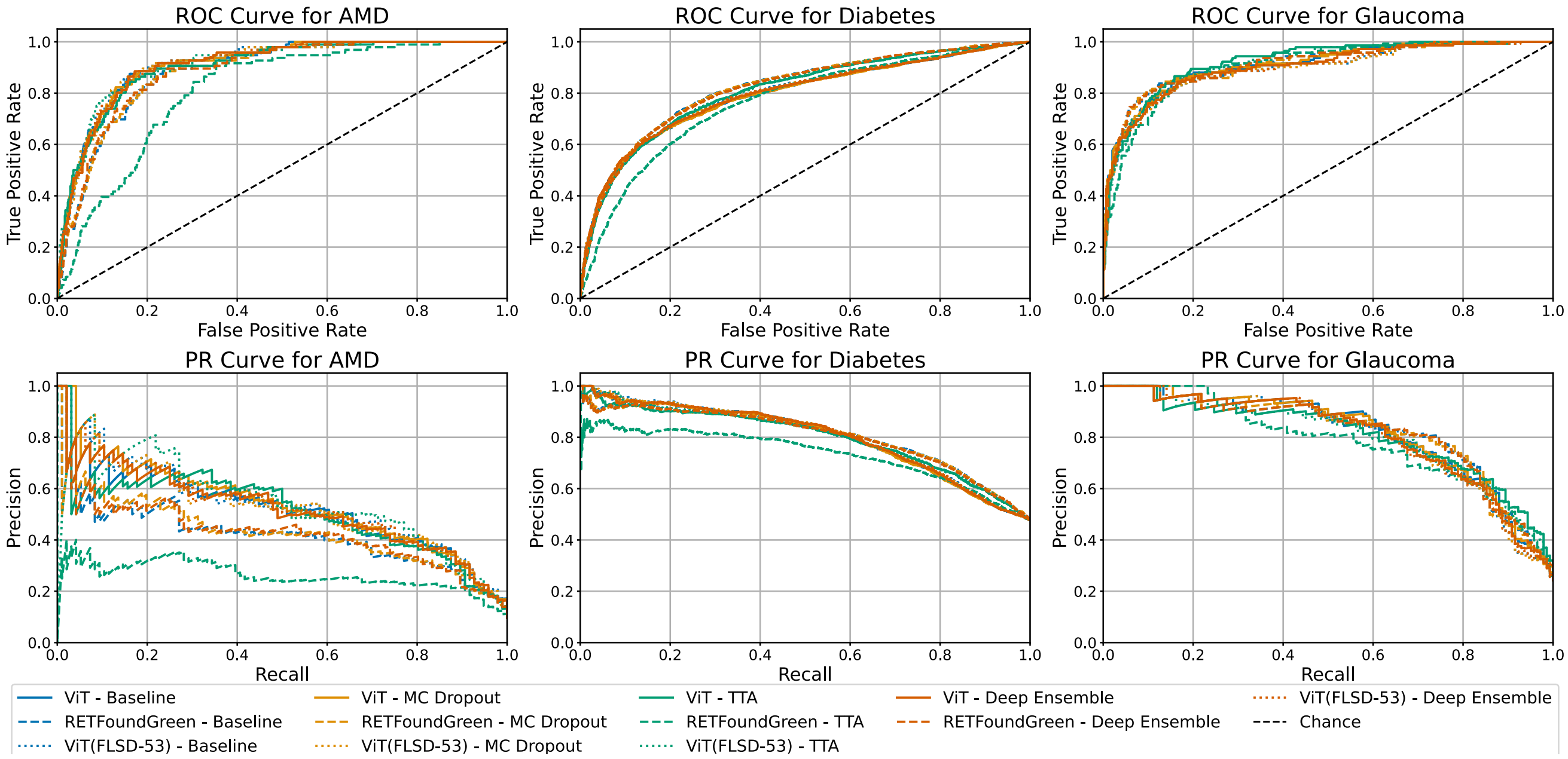

Figure 6. Larger version of Figure 1. Figures of ROC and PR curves for different trained models and inference strategies.

Table 5. Selective Prediction Benchmark Results evaluating AURC (↓), Risk@90% Coverage (↓), and Coverage@5% Risk (↑) across UQ methods for `ViT` (top), `RETFound-Green` (middle), and `ViT(FLSD-53)` (bottom).

| Disease | Baseline | MSP | Predictive Entropy | MC Dropout | TTA | Deep Ensemble Total | Aleatoric | Epistemic | Std |
|---|---|---|---|---|---|---|---|---|---|
| **AURC** | | | | | | | | | |
| AMD | 0.407 [0.367, 0.448] | 0.126 [0.102, 0.153] | 0.126 [0.102, 0.153] | 0.128 [0.103, 0.156] | 0.052 [0.041, 0.065] | 0.142 [0.116, 0.172] | 0.142 [0.116, 0.172] | 0.172 [0.144, 0.206] | 0.154 [0.128, 0.186] |
| Diabetes | 0.211 [0.199, 0.223] | 0.167 [0.154, 0.180] | 0.167 [0.154, 0.180] | 0.178 [0.165, 0.192] | 0.184 [0.171, 0.197] | 0.165 [0.152, 0.178] | 0.165 [0.152, 0.178] | 0.216 [0.201, 0.232] | 0.190 [0.177, 0.204] |
| Glaucoma | 0.224 [0.185, 0.260] | 0.078 [0.054, 0.104] | 0.078 [0.054, 0.104] | 0.078 [0.054, 0.104] | 0.042 [0.030, 0.055] | 0.078 [0.054, 0.103] | 0.078 [0.054, 0.103] | 0.132 [0.097, 0.167] | 0.106 [0.077, 0.135] |
| **Risk @ 90% Cov** | | | | | | | | | |
| AMD | 0.232 [0.205, 0.260] | 0.186 [0.160, 0.212] | 0.186 [0.160, 0.212] | 0.193 [0.167, 0.221] | 0.122 [0.100, 0.146] | 0.197 [0.169, 0.222] | 0.194 [0.169, 0.222] | 0.210 [0.184, 0.235] | 0.207 [0.182, 0.234] |
| Diabetes | 0.270 [0.257, 0.284] | 0.245 [0.231, 0.257] | 0.245 [0.231, 0.257] | 0.258 [0.246, 0.272] | 0.250 [0.238, 0.264] | 0.242 [0.229, 0.256] | 0.242 [0.229, 0.255] | 0.257 [0.243, 0.269] | 0.252 [0.239, 0.266] |
| Glaucoma | 0.186 [0.152, 0.221] | 0.145 [0.112, 0.174] | 0.145 [0.112, 0.174] | 0.146 [0.115, 0.178] | 0.108 [0.084, 0.135] | 0.141 [0.110, 0.171] | 0.140 [0.110, 0.170] | 0.152 [0.120, 0.183] | 0.144 [0.113, 0.175] |
| **Cov @ 5% Risk** | | | | | | | | | |
| AMD | 0.002 [0.000, 0.011] | 0.047 [0.015, 0.195] | 0.047 [0.015, 0.195] | 0.023 [0.012, 0.176] | 0.573 [0.476, 0.654] | 0.021 [0.007, 0.143] | 0.021 [0.007, 0.143] | 0.021 [0.004, 0.053] | 0.022 [0.012, 0.089] |
| Diabetes | 0.040 [0.014, 0.105] | 0.036 [0.014, 0.101] | 0.036 [0.014, 0.101] | 0.053 [0.020, 0.085] | 0.016 [0.005, 0.056] | 0.040 [0.014, 0.104] | 0.040 [0.014, 0.104] | 0.005 [0.001, 0.019] | 0.023 [0.010, 0.074] |
| Glaucoma | 0.074 [0.019, 0.147] | 0.437 [0.039, 0.558] | 0.437 [0.039, 0.558] | 0.360 [0.031, 0.625] | 0.608 [0.457, 0.762] | 0.385 [0.041, 0.570] | 0.385 [0.041, 0.575] | 0.038 [0.017, 0.218] | 0.175 [0.026, 0.406] |

| Disease | Baseline | MSP | Predictive Entropy | MC Dropout | TTA | Deep Ensemble Total | Aleatoric | Epistemic | Std |
|---|---|---|---|---|---|---|---|---|---|
| **AURC** | | | | | | | | | |
| AMD | 0.447 [0.407, 0.488] | 0.162 [0.132, 0.194] | 0.162 [0.132, 0.194] | 0.152 [0.123, 0.181] | 0.072 [0.058, 0.085] | 0.172 [0.142, 0.206] | 0.172 [0.141, 0.206] | 0.195 [0.160, 0.227] | 0.181 [0.149, 0.213] |
| Diabetes | 0.210 [0.198, 0.224] | 0.153 [0.141, 0.166] | 0.153 [0.141, 0.166] | 0.166 [0.154, 0.179] | 0.236 [0.220, 0.253] | 0.155 [0.143, 0.168] | 0.155 [0.143, 0.168] | 0.212 [0.198, 0.228] | 0.194 [0.181, 0.209] |
| Glaucoma | 0.171 [0.135, 0.211] | 0.055 [0.037, 0.076] | 0.055 [0.037, 0.076] | 0.056 [0.038, 0.076] | 0.053 [0.041, 0.065] | 0.057 [0.038, 0.079] | 0.057 [0.038, 0.079] | 0.090 [0.062, 0.120] | 0.071 [0.050, 0.096] |
| **Risk @ 90% Cov** | | | | | | | | | |
| AMD | 0.252 [0.224, 0.283] | 0.210 [0.183, 0.238] | 0.210 [0.183, 0.238] | 0.210 [0.184, 0.239] | 0.142 [0.122, 0.163] | 0.220 [0.193, 0.248] | 0.219 [0.194, 0.248] | 0.233 [0.206, 0.261] | 0.230 [0.203, 0.259] |
| Diabetes | 0.258 [0.246, 0.271] | 0.222 [0.210, 0.235] | 0.222 [0.210, 0.235] | 0.236 [0.224, 0.249] | 0.286 [0.274, 0.300] | 0.225 [0.213, 0.238] | 0.224 [0.212, 0.238] | 0.252 [0.239, 0.266] | 0.243 [0.230, 0.257] |
| Glaucoma | 0.139 [0.110, 0.167] | 0.098 [0.072, 0.124] | 0.098 [0.072, 0.124] | 0.104 [0.081, 0.133] | 0.134 [0.111, 0.156] | 0.100 [0.076, 0.126] | 0.101 [0.076, 0.126] | 0.116 [0.093, 0.148] | 0.116 [0.089, 0.145] |
| **Cov @ 5% Risk** | | | | | | | | | |
| AMD | 0.001 [0.000, 0.007] | 0.003 [0.001, 0.123] | 0.003 [0.001, 0.123] | 0.014 [0.008, 0.149] | 0.393 [0.325, 0.488] | 0.004 [0.001, 0.088] | 0.004 [0.001, 0.088] | 0.000 [0.000, 0.025] | 0.004 [0.001, 0.067] |
| Diabetes | 0.011 [0.002, 0.082] | 0.011 [0.002, 0.082] | 0.011 [0.002, 0.082] | 0.014 [0.004, 0.069] | 0.001 [0.000, 0.007] | 0.009 [0.002, 0.074] | 0.009 [0.002, 0.074] | 0.005 [0.000, 0.052] | 0.029 [0.000, 0.070] |
| Glaucoma | 0.055 [0.026, 0.140] | 0.497 [0.200, 0.760] | 0.497 [0.200, 0.760] | 0.515 [0.260, 0.735] | 0.589 [0.527, 0.704] | 0.517 [0.101, 0.738] | 0.519 [0.101, 0.740] | 0.108 [0.031, 0.421] | 0.317 [0.058, 0.615] |

| Disease | Baseline | MSP | Predictive Entropy | MC Dropout | TTA | Deep Ensemble Total | Aleatoric | Epistemic | Std |
|---|---|---|---|---|---|---|---|---|---|
| **AURC** | | | | | | | | | |
| AMD | 0.402 [0.364, 0.438] | 0.122 [0.099, 0.147] | 0.122 [0.099, 0.147] | 0.168 [0.141, 0.196] | 0.132 [0.105, 0.161] | 0.126 [0.102, 0.151] | 0.126 [0.103, 0.152] | 0.171 [0.141, 0.200] | 0.165 [0.137, 0.194] |
| Diabetes | 0.211 [0.199, 0.224] | 0.163 [0.151, 0.175] | 0.163 [0.151, 0.175] | 0.250 [0.233, 0.268] | 0.267 [0.250, 0.285] | 0.164 [0.152, 0.177] | 0.164 [0.152, 0.177] | 0.274 [0.257, 0.292] | 0.269 [0.252, 0.287] |
| Glaucoma | 0.209 [0.172, 0.250] | 0.072 [0.052, 0.096] | 0.072 [0.052, 0.096] | 0.120 [0.086, 0.161] | 0.116 [0.084, 0.154] | 0.081 [0.059, 0.106] | 0.081 [0.060, 0.106] | 0.177 [0.133, 0.224] | 0.171 [0.128, 0.219] |
| **Risk @ 90% Cov** | | | | | | | | | |
| AMD | 0.230 [0.203, 0.257] | 0.179 [0.152, 0.203] | 0.179 [0.152, 0.203] | 0.195 [0.170, 0.221] | 0.179 [0.155, 0.206] | 0.182 [0.158, 0.207] | 0.187 [0.162, 0.211] | 0.203 [0.178, 0.230] | 0.201 [0.176, 0.226] |
| Diabetes | 0.272 [0.259, 0.285] | 0.238 [0.225, 0.251] | 0.238 [0.225, 0.251] | 0.264 [0.251, 0.277] | 0.262 [0.248, 0.275] | 0.239 [0.227, 0.252] | 0.239 [0.227, 0.251] | 0.265 [0.253, 0.278] | 0.263 [0.251, 0.276] |
| Glaucoma | 0.173 [0.141, 0.209] | 0.132 [0.106, 0.165] | 0.132 [0.106, 0.165] | 0.137 [0.110, 0.170] | 0.129 [0.103, 0.159] | 0.139 [0.113, 0.173] | 0.143 [0.114, 0.174] | 0.164 [0.131, 0.194] | 0.157 [0.127, 0.189] |
| **Cov @ 5% Risk** | | | | | | | | | |
| AMD | 0.003 [0.000, 0.013] | 0.060 [0.024, 0.156] | 0.060 [0.024, 0.156] | 0.003 [0.000, 0.094] | 0.013 [0.007, 0.087] | 0.050 [0.015, 0.170] | 0.050 [0.014, 0.171] | 0.014 [0.008, 0.084] | 0.021 [0.008, 0.085] |
| Diabetes | 0.060 [0.034, 0.116] | 0.052 [0.034, 0.117] | 0.052 [0.034, 0.117] | 0.001 [0.000, 0.003] | 0.000 [0.000, 0.007] | 0.042 [0.017, 0.093] | 0.042 [0.017, 0.092] | 0.000 [0.000, 0.001] | 0.000 [0.000, 0.001] |
| Glaucoma | 0.089 [0.022, 0.140] | 0.394 [0.223, 0.611] | 0.394 [0.223, 0.611] | 0.003 [0.000, 0.194] | 0.038 [0.009, 0.176] | 0.313 [0.070, 0.531] | 0.313 [0.074, 0.531] | 0.002 [0.000, 0.039] | 0.002 [0.000, 0.039] |

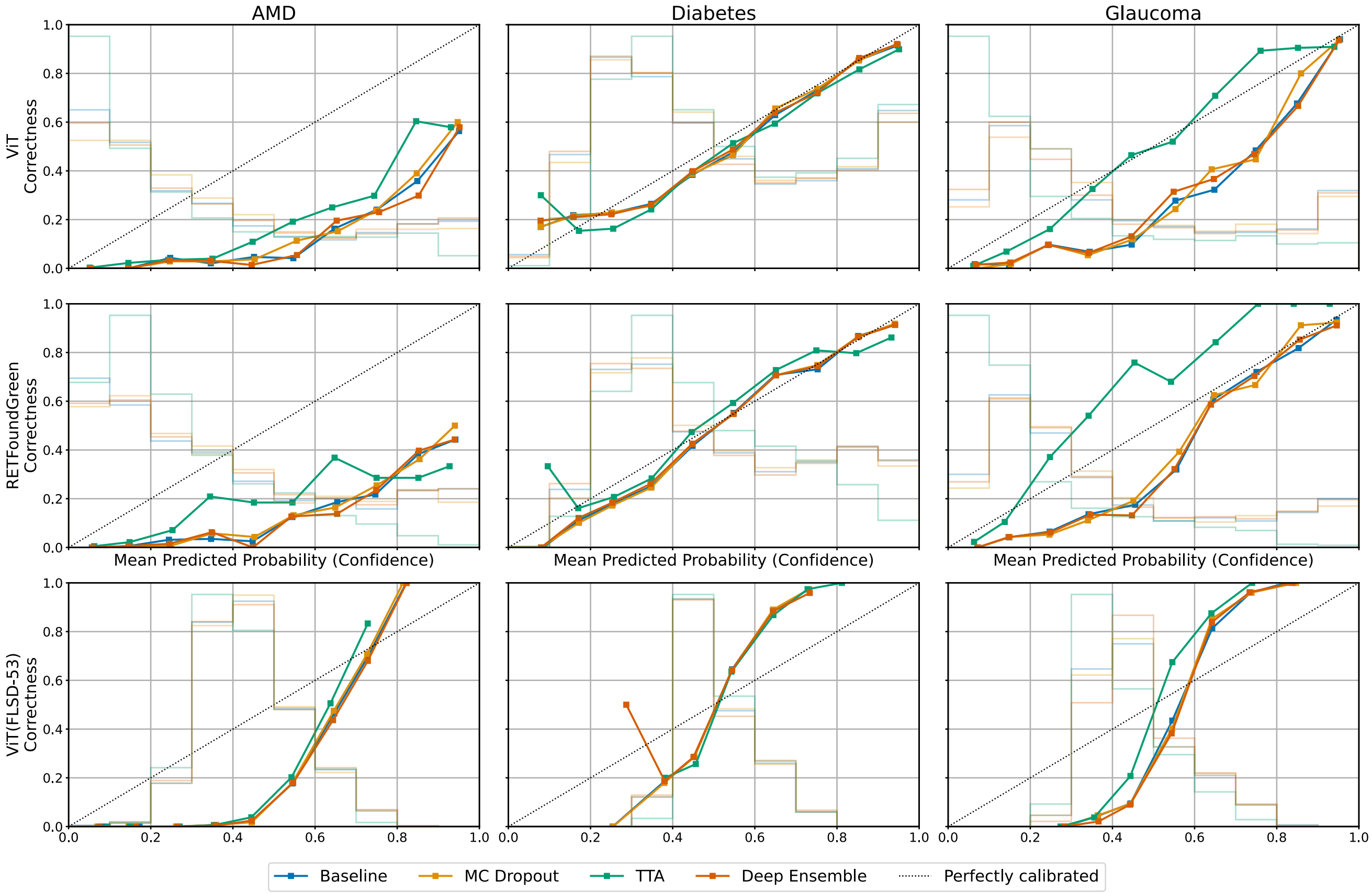

Figure 7. Larger version of Figure 2. Reliability diagrams for disease detection models, where well-calibrated methods lie near the diagonal.

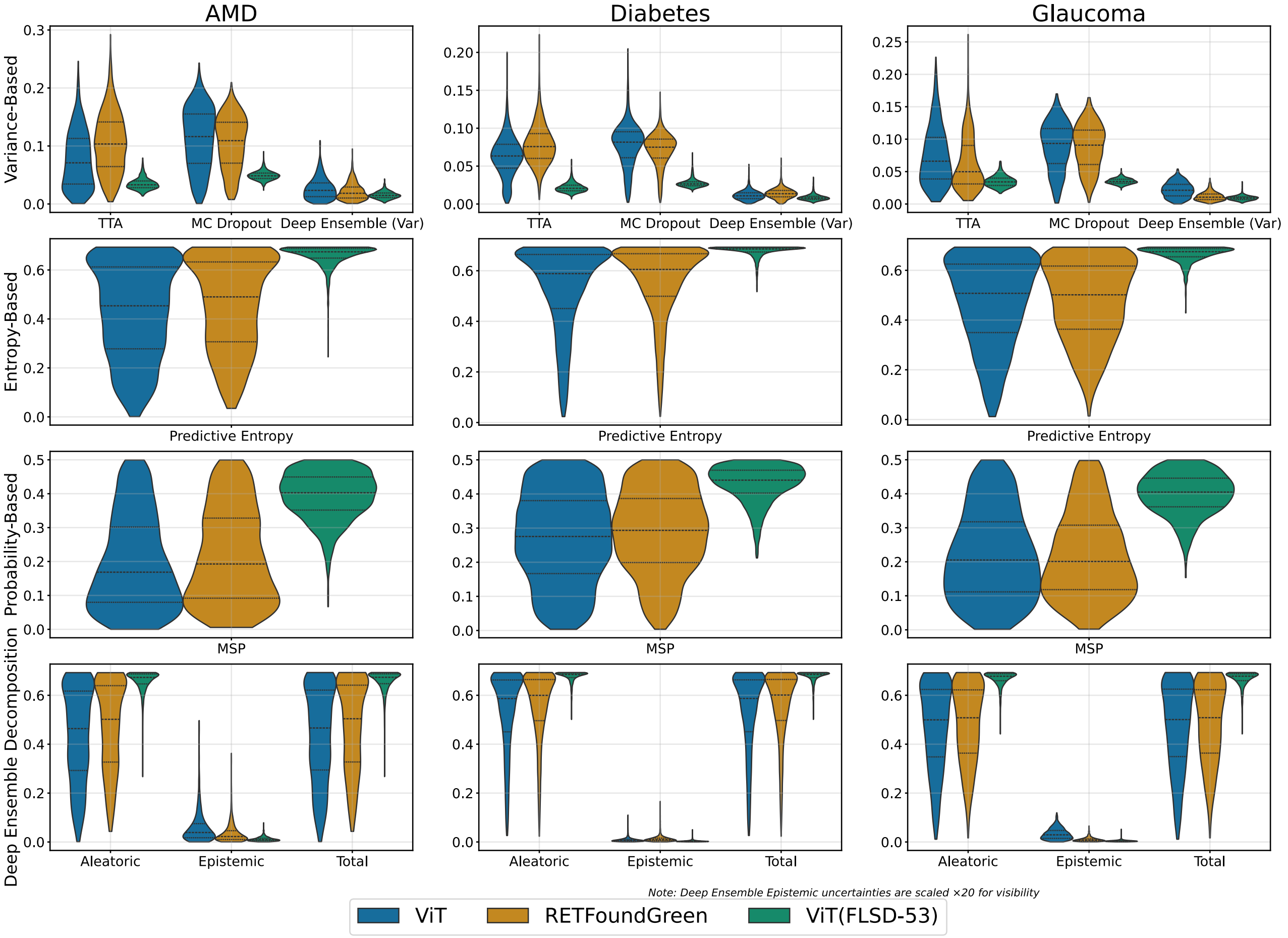

Figure 8. Violin distributions illustrating the diversity and spread of normalized uncertainty scores across retinal diseases and UQ methods.

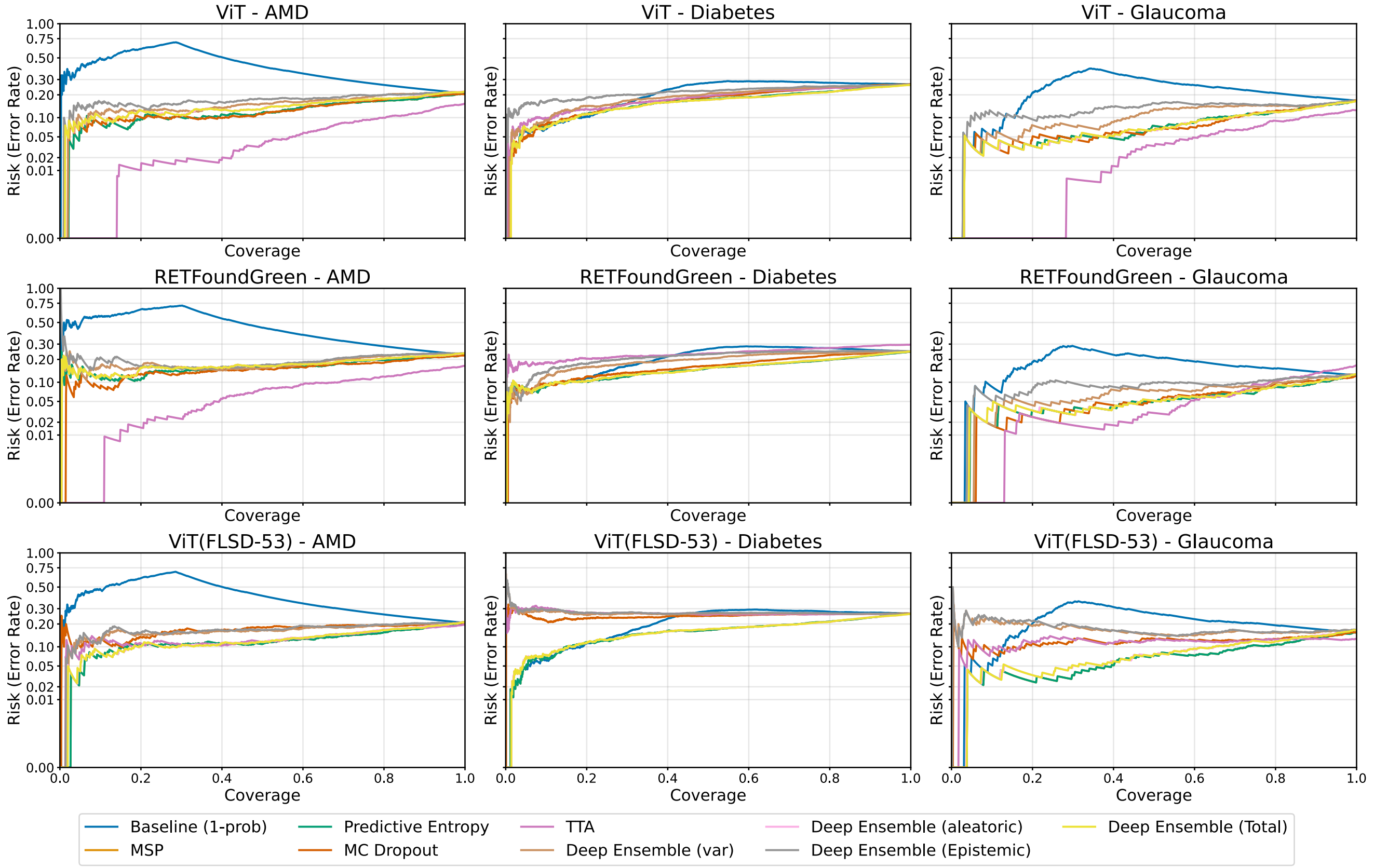

Figure 9. Larger version of Figure 3. Figures of Risk–Coverage analysis showing the effectiveness of UQ methods. The x-axis shows coverage, the proportion of test instances the system predicts automatically, with the rest rejected (abstained/referred) for expert review. For example, at 90% coverage, the 10% of test samples with the highest uncertainty scores are removed and referred to an expert.

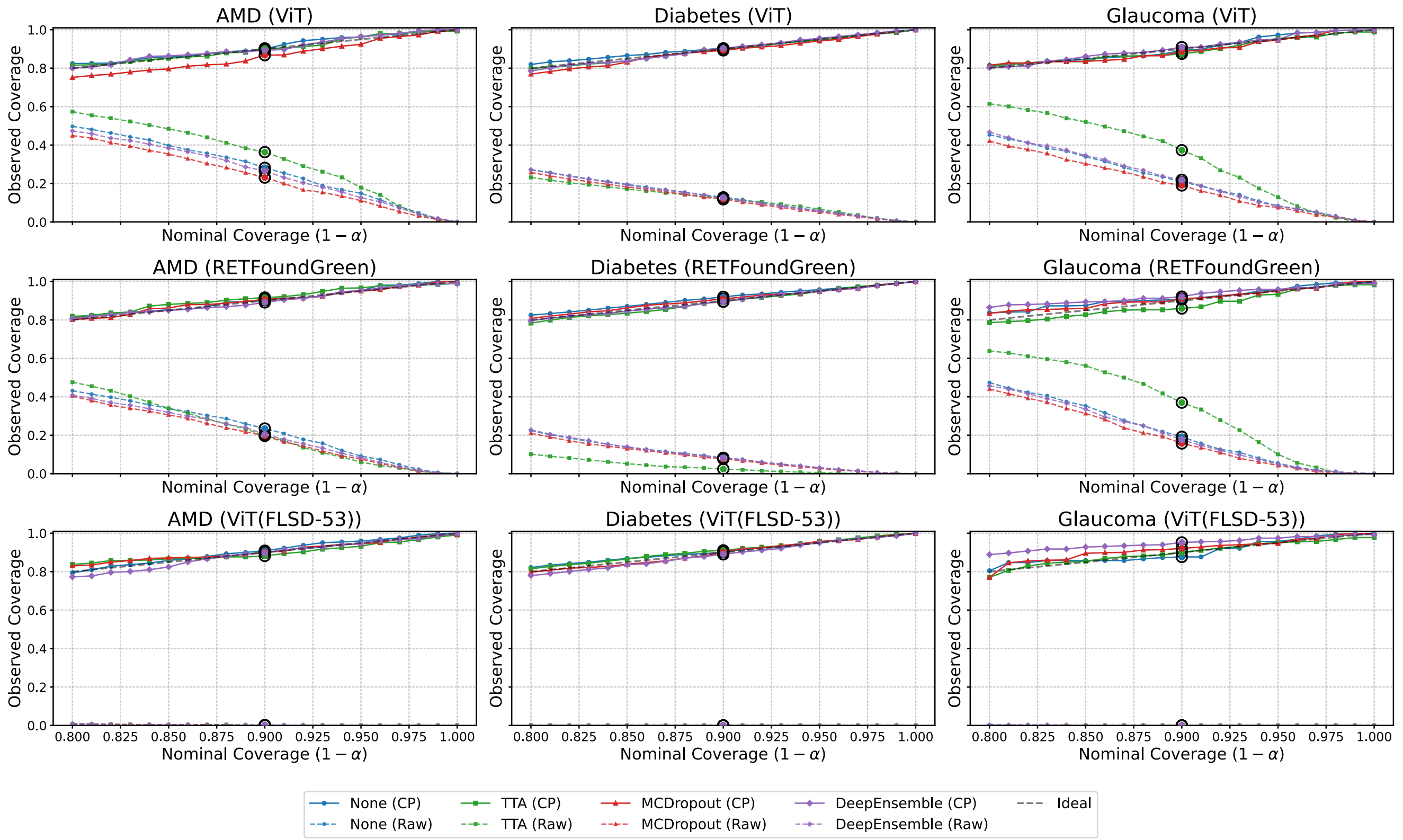

Figure 10. Larger version of Figure 4. Observed vs. nominal coverage showing under-coverage in raw predictions and correction with CP. All the raw methods exhibit undercoverage and overconfidence, whereas CP effectively compensates for these shortcomings.

Table 6. Observed coverage with and without conformal prediction (coverage 90%) across diseases and methods.

| | None | MC Dropout | Deep Ensemble | TTA |
|---|---|---|---|---|
| **Observed Coverage (CP) (%)** | | | | |
| AMD | 89.91 | 86.78 | 89.30 | 90.31 |
| Diabetes | 90.31 | 89.27 | 90.27 | 90.18 |
| Glaucoma | 89.21 | 88.70 | 90.92 | 87.50 |
| **Observed Coverage (Raw) (%)** | | | | |
| AMD | 28.15 | 23.11 | 26.54 | 36.33 |
| Diabetes | 12.82 | 11.80 | 12.52 | 12.27 |
| Glaucoma | 20.72 | 19.01 | 21.92 | 37.33 |

ViT

| | None | MC Dropout | Deep Ensemble | TTA |
|---|---|---|---|---|
| **Observed Coverage (CP) (%)** | | | | |
| AMD | 90.21 | 90.72 | 89.10 | 91.62 |
| Diabetes | 91.99 | 90.99 | 89.46 | 89.63 |
| Glaucoma | 90.58 | 91.10 | 92.12 | 85.96 |
| **Observed Coverage (Raw) (%)** | | | | |
| AMD | 23.51 | 19.68 | 20.48 | 19.88 |
| Diabetes | 8.26 | 7.69 | 8.16 | 2.43 |
| Glaucoma | 19.18 | 15.75 | 17.64 | 36.99 |

RETFoundGreen

| | None | MC Dropout | Deep Ensemble | TTA |
|---|---|---|---|---|
| **Observed Coverage (CP) (%)** | | | | |
| AMD | 90.92 | 90.21 | 89.71 | 88.19 |
| Diabetes | 90.10 | 90.20 | 89.01 | 91.08 |
| Glaucoma | 87.67 | 92.12 | 95.21 | 90.07 |
| **Observed Coverage (Raw) (%)** | | | | |
| AMD | 0.20 | 0.10 | 0.10 | 0.10 |
| Diabetes | 0.00 | 0.00 | 0.00 | 0.00 |
| Glaucoma | 0.00 | 0.00 | 0.00 | 0.00 |

ViT(FLSD-53)